\documentstyle[floats,twocolumn,prl,aps]{revtex}
\begin{document}
\draft
\def\ref{\par\noindent\hangindent=3mm\hangafter=1}
\narrowtext
{
\title{
The Boson Mediators of High-$T_c$ Superconductivity:
Phonons Versus Composite Bosons from the Superconducting
Phenomenology}
}
\author{Georgios Varelogiannis and Luciano Pietronero}
\address{
Dipartimento di Fisica,
Universita di Roma ``La Sapienza'', Piazzale Aldo Moro 2, I-00185 Roma,
Italy}
\author{\parbox{397pt}{\vglue 0.3cm \small
We address the question of whether boson mediators of high-$T_c$
superconductivity are composite (electronic) or independent phonons.
For s-wave superconductivity we show from the available experiments
that the hypothesis of composite bosons is rather unlikely.
Our analysis points naturally towards phonon mediators.
In addition we point out that
the eventual presence of a peak in the temperature
dependence of the microwave conductivity while the Hebel-Slichter
peak is absent in the temperature dependence of the NMR relaxation
rate, can be understood within a phonon mechanism if one takes
into account the modulation of the electron-phonon
coupling (predominance of forward scattering)
induced by Coulomb correlation of the carriers. }}
\maketitle
\par
The discovery of high temperature superconductivity has
stimulated important theoretical activity
\cite{Gre}. There is no consensus on the pairing
mechanism responsible for high-$T_c$ superconductivity.
Motivated by the phenomenology of superconducting
materials in the normal state many
new ideas on the pairing mechanism have been proposed.
Various authors have conjectured that the pairing interaction
could result purely from the strong Coulomb correlations of the
carriers
\cite{Dagg}.
Independent of the
considered microscopic model in the normal state,
{\it a boson mediated interaction is necessary}
\cite{nambu}
for the dynamic symmetry breaking associated with a superconducting
transition.
In conventional theories the role of the boson mediators is
played by phonons. In the new models proposed for high-$T_c$
superconductivity, the role of the boson mediators should be played either
by antiferromagnetic paramagnons \cite{Pines}, by composite excitons
such as slave bosons \cite{Varma} or even by gauge bosons
\cite{Lee}.
In all cases
an Eliashberg-like approach is necessary
for the description of the superconducting
state.
Eliashberg theory (ET) \cite{Elias}
might be viewed as an interface between microscopic
theories and superconducting phenomenology.
In general it is hard to learn microscopic information about
the nature of the coupling from the superconducting phenomenology.
In a few cases however this is possible and it represents very
important information because it relates directly to
the physics relevant for superconductivity.

Our discussion will be limited essentially to the case of s-wave pairing.
However our perspective is that a number of experimental
anisotropic effects in the high-$T_c$ materials, just reflect the anisotropies
of the electronic density of states because of the predominance of
small $q$ scattering in the isotropic s-wave interaction \cite{PRLsub}
We treat here the question of whether {\it the
bosons which mediate
the pairing interaction are bosons such as phonons independent
from the carriers or
some composite excitons such as slave bosons}.
This represents a crucial theoretical question
in high-temperature superconductivity.
The relevant points
of the phenomenology
merit in our opinion an intensive investigation analogous to that
underway for the determination of the symmetry of the pairing.
Our basic point is that, in the case of a composite boson,
a gap must open in this boson spectrum below $T_c$ while this is not
the case for phonons. This low-energy cutoff in the boson spectrum allows
us to distinguish between the two possibilities. Clearly such an effect
will be evident for s-wave pairing while for d-wave pairing it mixes
with the intrinsic anisotropy.

The composite boson spectrum which mediates superconductivity
must
arise from the dynamical susceptibility
of the
fermionic system.
In this respect we are considering the situation in which the
electronic system that leads to the superconducting pairs is the same as
that which provides the composite bosons that mediate superconductivity.
The susceptibility can always be decomposed into a
sum of a bare susceptibility plus an
interacting one, as shown in figure 1.
The vertex $\Gamma$ depends on the mechanism
under consideration.
In the context of a Marginal Fermi Liquid (MFL) \cite{Kuroda}
a particular assumption is made for the frequency
dependence of $\Gamma$ motivated by the normal state
dynamical properties of the oxides.
The following discussion is more general
and remains valid independent of the precise structure of $\Gamma$
and of the nature of the underlying mechanism.

When the
s-wave superconducting state appears,
a superconducting gap $\Delta_g$
opens in all momentum directions,
eventually with a modulation in the amplitude due to anisotropic effects.
This corresponds to
the opening of a gap of $2\Delta_g$
in the dynamical susceptibility \cite{Kuroda},
since at zero temperature all electrons are paired and
one has to excite
a pair in order to create a fluctuation.
As a consequence
the onset of superconductivity implies
the opening of
a gap of $2\Delta_g(T)$ in any composite boson spectrum
(see Fig. 2).
On the other hand,
if superconductivity is mediated by
phonons
this gap of $2\Delta_g$
is absent. Therefore identifying the presence and the
magnitude of a gap in the boson
spectrum is a crucial point for the understanding of
the microscopic mechanism of high-$T_c$ superconductivity.

The relevant experimental tool for the investigation of this question
is infrared spectroscopy (IR). We can take advantage from the fact that
the coherence length of superconductivity $\xi$ in high-$T_c$
materials
is shorter than the mean free path $l$ for scattering
by non magnetic impurities or defects. In that case
the threshold energy for IR absorption at
$T\rightarrow 0$ directly reflects the
minimal energy of the eventual composite boson spectrum.
In fact, although in the
dirty limit ($\xi\gg l$), valid for all
conventional low-$T_c$ superconductors, the IR
absorption starts at an energy equal to $2\Delta_g$,
in the clean limit ($\xi < l$), relevant for
high-$T_c$ materials, absorption starts
at $2\Delta_g+\Omega_{min}$
(see Fig. 3),
where $\Delta_g$ is the superconducting gap and
$\Omega_{min}$ is the minimal energy of the boson spectrum.
Clearly for the case of a boson composite spectrum
$\Omega_{min}=2\Delta_g$ as we have mentioned previously.
This difference between clean and dirty limit
is due to the fact that, in the
usual dirty limit, impurities lead to a damping and a decay channel
for the electrons. In the clean limit, instead the only decay channel
available is via phonons and therefore the structure of the
phonon density of states is reflected in the IR spectrum.

In high-$T_c$ superconductors, many IR experiments
on different materials and samples report a feature which can be
associated with the threshold energy for absorption
at an energy $E_{IR}=2\Delta_{IR}\approx 8T_c$
\cite{irgap}. On the other hand from the one particle tunnel spectroscopies,
the reported gap values are
$2\Delta_g\approx 5.5T_c-6T_c$ \cite{batlog}. The minimal energy
of the boson spectrum
$\Omega_{min}$ thus
appears to be equal to $2-2.5T_c$ and {\it not}
equal to $2\Delta_g$ as
would be the case for a composite boson spectrum similar
to that
considered in the MFL framework \cite{carbo}.
This can be considered as {\it direct evidence for the exclusion
of the composite boson scenario in an s-wave approach}.

The fact that the boson spectrum has a low energy cut-off
of $\approx 2T_c-2.5T_c $ {\it supports} phonon mediated superconductivity.
The existence of this
cut-off means that the coupling
of electrons to phonons with energies lower than $2.5T_c$
is negligible compared to their coupling to the phonons with
energies larger than $2.5T_c$. There is a simple
physical explanation for this behavior. The low energy
part of the phonon spectrum is mainly associated with vibrations of
the heavy molecules far from the $CuO$ planes.
In the case of $YBa_2Cu_3O_7$ the spectral region
$\Omega\leq 3T_c$ concerns principally vibrations of $Y$ and $Ba$ \cite{ZZ}.
The in-plane electrons responsible for superconductivity
are far from those molecules and therefore the
coupling to these vibrations that could be relevant for
superconductivity is very small.
On the other hand at larger energies one can find the
optical $\vec{q}=0$ vibrations of the in-plane
oxygens \cite{liu}, for which there is experimental
evidence from Raman \cite{raman,ZZ} neutron \cite{neutron}
and recent site-selective isotope measurements \cite{Liarokapis}
that are strongly coupled to the carriers.
Note that
a systematic analysis of the
spectral dependence of the gap ratio, leads to the conclusion
that in high-$T_c$ cuprates the coupling might be concentrated
precisely in this
spectral region $25-50meV$ \cite{prbmoi}.

The discrepancy between
the gap values reported from infrared
($2\Delta_{IR}\approx 8T_c$) and the gap values
reported from other spectroscopies ($2\Delta_g\approx 5.5T_c-6T_c$)
is therefore associated with
the fact that
cuprates are in the ``clean'' regime ($\xi<l$).
As a consequence
$2\Delta_{IR}\approx 2\Delta_g+\Omega_{min}$ with
$\Omega_{min}\approx 2.5T_c$ in the phonon spectrum of cuprates
imposed by the strong two-dimensional character of their
superconductivity, that leads to a very small coupling to the low
frequency phonons.

Let us consider now the central argument
used to support the composite spectrum
s-wave scenario from the high-$T_c$ superconducting
phenomenology. In the NMR relaxation experiments
it is clear that no Hebel-Slichter peak is present
in the high-$T_c$ superconductors. This does
not contradict
conventional theories since for
the strong couplings
necessary to understand the
gap ratio values of those materials \cite{prbmoi}
and the dip-like structure in the density of states of excitations
\cite{prbmoiRC},
the Hebel-Slichter peak is consistently absent
\cite{AR}. This is understood
as a quasiparticle lifetime effect
and it is qualitatively and
quantitatively consistent within Eliashberg theory with
the rest of the high-$T_c$ phenomenology \cite{moiPhC,moiPLA,russes}.
However
in some microwave spectroscopy experiments
a peak structure is reported in the temperature
dependence of the ratio of the real part of the conductivity
in the superconducting state over that in the
normal state $\sigma_{1s}/\sigma_{1n}$
\cite{Nuss,Bonn,Holczer}.

These
data seem to contradict
conventional phonon mediated superconductivity. In fact
the absence of coherence effects in the NMR relaxation should exclude
a peak structure in the temperature dependence of the microwave
conductivity \cite{moiPLA}.
This contradiction disappears if one supposes
that the superconductivity is due to some composite
bosons arising from some internal interaction of the
carriers such as strong Coulomb correlations. In fact the
peak structures in the
temperature dependence of the microwave
conductivity can be understood as arising from a strong reduction
of the scattering rate just below $T_c$ \cite{Nuss,Bonn}. This strong
reduction arises naturally with the opening of the $2\Delta_g$
gap in the spectrum of the composite bosons \cite{Nuss,carbo}.
In fact at the onset of superconducting transition the
$\Omega\leq 2\Delta_g$ part of the composite spectrum disappears
and the diffusion of the electrons
with this part of the spectrum is frozen \cite{Nuss}
(see Fig. 2). The data of Refs. \cite{Nuss,Bonn} have therefore been
considered as evidence that the boson mediators are composite and that
superconductivity is due primarily to electronic mechanisms \cite{Nuss}.

However we are going to see that the presence of a peak
in the microwave conductivity is actually
{\it not contradictory to phonon mediated superconductivity}.
We studied
the effect of strong Coulomb correlations on the electron-phonon
coupling \cite{VarPre}. The main conclusion of our
studies, in agreement with previous results \cite{Abrik}, is that
strong Coulomb correlations can lead to a ``modulation''
in the momentum dependence
of the electron-phonon interaction, enhancing the small momentum
transfer
processes and reducing the large momentum transfer processes.
In particular
this results in an enhancement of the
coupling of electrons with
$q/k_F\ll 1$ phonons.
Such a momentum modulation indicates the vicinity of
a phase separation instability in the electronic system \cite{Crete}.
However, after the appearance of
phonon mediated superconducting order
the effect of the Coulomb correlations
is renormalized by retardation effects
\cite{VarPre}.
In fact all condensed electrons are
correlated to be not closer than a characteristic distance related
to the characteristic time for the absorption of the virtual phonon.

As a consequence,
in the superconducting state
the ``modulation'' in the momentum dependence
of the electron-phonon interaction
induced by the Coulomb correlations will
be reduced \cite{VarPre} and this
results in a drop of the $\vec{q}\rightarrow 0$
part of the electron-phonon interaction (see Fig. 4).
But the microwave spectroscopies are particularly
sensitive to the $\vec{q}\rightarrow 0$ processes since
the scattering rate is strongly influenced by these processes.
The effect on the microwave conductivity
of this reduction of the $\vec{q}\rightarrow 0$
part of the electron-phonon coupling
is completely analogous to the effect of the
{\it opening} of the $2\Delta_g$ gap in a composite boson spectrum
\cite{VarPre}.
On the other hand, for the NMR relaxation the relevant processes
are local, and therefore they correspond to the sum over all momenta of the
dynamic susceptibility. In this manner, the reduction of the
``modulation'' in the momentum dependence of
the electron-phonon coupling below the superconducting
transition does not have any influence on the temperature
dependence of the NMR relaxation rate \cite{VarPre}.
The difference in the behaviors of the NMR relaxation rate and
microwave conductivity is thus understood as arising from the
fact that the first-one {\it is a local experiment while the second one is
a longwavelength experiment}.

In conclusion we have shown that the present superconductivity
phenomenology seems to support a picture based on s-wave and electron-phonon
coupling dominated by small $q$ processes.
The presence of a peak in the temperature dependence of the
microwave conductivity while a Hebel-Slichter peak is absent in the
NMR relaxation rate, {\it cannot} be considered as an evidence
for a purely electronic mechanism since it can also be understood in the
context of small $q$ phonon mediated superconductivity.
Notice that
the enhancement of
forward electron-phonon scattering
turns out to be
favorable for superconductivity
if one takes into account non adiabatic corrections in the
electron-phonon vertex \cite{Pietro},
and could also give some insight on the origin of
large couplings in an Eliashberg description
of the high-$T_c$ phenomenology \cite{lpmoi}.

We acknowledge interesting discussions with
P. Lederer, O. Dolgov,
C. Castellani, B. Chakraverty, C. DiCastro, M. Grilli and R. Zeyher.
GV is supported by the European Union
under contract ERBCHBICT930906.



{\Large \bf Figure Captions}

{\bf Figure 1:} The susceptibility of the strongly correlated
electronic system decomposed in the sum of a bare susceptibility
plus an interacting one.
In a scenario of superconductivity based on a purely electronic
mechanism, the boson mediators of the pairing interaction
are proportional to this susceptibility.

{\bf Figure 2:} The Eliashberg function for a purely electronic mechanism
of superconductivity.
When the superconducting gap $\Delta_g$ opens, a gap
of $2\Delta_g$ opens in this Eliashberg spectrum.
Since the shaded area of the
spectrum disappears, the scattering of the carriers with that part
of the spectrum is frozen and this results to a drop of the
scattering time for small momenta.

{\bf Figure 3:}
The real part of the infrared conductivity for different schemes.
Full line corresponds to the dirty limit $\xi\gg l$ of the
electron-phonon coupling spectrum shown in the inset. Absorption
starts at an energy equal to $2\Delta_g$ (here $\lambda=2.7$ and
$\Delta_g=0.65$ in the
frequency units of the spectrum). The dot-dashed line corresponds
to the clean limit $\xi\ll l$ of the same electron-phonon spectrum
with the same $\lambda$ and the same $\Delta_g$.
In the clean limit absorption starts at a higher energy
$2\Delta_g+\Omega_{min}$ (as can be seen in the inset, where $\Omega_{min}=0.7$
in the frequency units of the spectrum). The dotted line corresponds to
to the clean limit of
a composite exciton spectrum as that shown in figure 2 with the
same $\Delta_g$ as previously.
In this last case absorption starts at $4\Delta_g$.

{\bf Figure 4:} The momentum dependence of the
electron-phonon scattering amplitude \cite{VarPre}.
In the normal state, strong correlations
enhance small momentum transfer processes and reduce large
momentum transfer processes. In the phonon mediated
superconducting state the effect of correlations is
renormalized because of retardation and this results in a drop
of the scattering time for small momenta, that can be analogous to that
produced by the opening of the $2\Delta_g$ gap
in the Eliashberg function of a purely electronic mechanism (see Fig. 2).

\end{document}